\def\la{\hbox{{\lower -2.5pt\hbox{$<$}}\hskip -8pt\raise
-2.5pt\hbox{$\sim$}}}
\def\ga{\hbox{{\lower -2.5pt\hbox{$>$}}\hskip -8pt\raise
-2.5pt\hbox{$\sim$}}}

\documentstyle{elsart}

\begin{document}
\begin{frontmatter}
\title{ULTRA HIGH ENERGY COSMIC RAYS: the theoretical challenge }
\author{A. V. Olinto\thanksref{corr}}
\thanks[corr]{Corresponding author. E-mail: olinto@oddjob.uchicago.edu}
\address{Department of Astronomy \& Astrophysics, \\ \& Enrico Fermi
Institute, \\
The University of Chicago, Chicago, IL 60637}
 
\begin{abstract}
The origin of the highest-energy cosmic rays remains a mystery. 
The lack of a high energy cutoff in the cosmic ray spectrum  
together with an apparently isotropic distribution of arrival
directions have strongly constrained most models proposed  for the
generation of these particles. An overview of the present state of
theoretical proposals is presented.
Astrophysical accelerators as well as top-down scenarios are reviewed
along with their   most general signatures.   The origin and nature of
these ultra-high energy particles will be tested by future
observations and may indicate as well as constrain physics beyond the
standard model of particle physics.

\end{abstract}

\end{frontmatter}

\section{Introduction}

The  detection of cosmic rays with energies above $10^{20}$ eV  has
triggered considerable interest on  the  origin and nature of these
particles. As reviewed by Watson \cite{watson99} in this volume, many
hundreds of events with energies above
$10^{19}$ eV and about 20 events above $10^{20}$ eV  have now been
observed by a number of experiments such as AGASA
\cite{takeda98,takeda99,Haya94}, Fly's Eye
\cite{bird},  Haverah Park \cite{L91}, Yakutsk \cite{Yak90}, and most
recently the High Resolution Fly's Eye \cite{hires}. 

Most unexpected is the significant flux of
events observed above $\sim 7 \times 10^{19}$ eV \cite{takeda98}  with no
sign of the Greisen-Zatsepin-Kuzmin (GZK) cutoff \cite{GZK66}. 
A cutoff should be present if the  ultra-high energy
particles are protons, nuclei, or photons from  extragalactic sources. 
Cosmic ray protons of energies above  a few $10^{19}$ eV lose energy
to photopion production off the cosmic microwave background (CMB) and
cannot originate further than about $50\,$Mpc away from Earth. Nuclei 
are photodisintegrated on shorter distances due to the infrared
background \cite{PSB76SS99} while the radio background
constrains photons to originate from even closer systems 
\cite{bere70PB96}.

In addition to the presence of events past the GZK cutoff, there has
been no clear counterparts identified in the  arrival direction of the
highest energy events. If these events are protons,  cosmic ray
observations should finally become astronomy!   At these high energies
the Galactic and extragalactic magnetic fields do not affect their
orbits significantly so that they should point
back to their sources within a few degrees. Protons at $10^{20}$
eV propagate mainly in straight lines as they traverse the Galaxy since
their gyroradii are $\sim $ 100 kpc in $ \mu$G  fields which is
typical in the Galactic disk. Extragalactic fields are expected to
be $\ll \mu$G \cite{KronVallee,BBO99}, and induce at most  
$\sim$ 1$^o$ deviation from the source. Even if
the Local Supercluster has relatively strong fields, the highest energy
events may deviate at most $\sim$ 10$^o$  
\cite{RKB98,SLB99}.  At present, no correlations between arrival directions
and plausible optical counterparts  such as sources in the Galactic
plane, the Local Group, or the Local Supercluster have been clearly
identified.  Ultra high energy cosmic ray (UHECR) data are consistent
with an isotropic distribution of sources in sharp contrast to the
anisotropic distribution of light within 50 Mpc from Earth.

The absence of a GZK cutoff and the isotropy of arrival directions are
two of the many challenges that models for the origin of UHECRs face.
This is an exciting open field, with many scenarios being proposed but 
no clear front runner.  Not only the origin of these 
particles  may be due to physics beyond the standard model of particle
physics, but their existence  can be used to constrain extensions of the
standard model such as violations of Lorentz invariance (see, e.g.,
\cite{ABGG00}). 

In the next section, a brief summary of the
challenges faced by all theoretical models is given. 
In \S3, astrophysical accelerators or ``bottom-up'' scenarios are
reviewed, hybrid models are discussed in \S4, and
top-down scenarios  in \S5.  To conclude,
future observational tests of UHECR models and their implications are
discussed  in \S6. For previous reviews of UHECR models, the reader
is encouraged to consult
\cite{hillas84,BBDGP90,gaisser90,bland99,berez99,BS99}. 

\section{The Challenge}

In attempting to explain the origin of UHECRs, models confront a number
of challenges. The extreme energy is the greatest challenge that models
of astrophysical acceleration face while for top-down models the
observed flux represents the highest hurdle. To complete the puzzle,
models have to match the spectral shape, the primary composition, and
the arrival direction distribution of the observed events.

{\it 2.1 Energy}

The observed  highest energy event at $3.2 \times 10^{20}$ eV
\cite{bird} argues for the existence of Zevatrons in
nature \cite{bland99}, accelerators that reach as high as one ZeV 
(ZeV=10$^{21}$ eV) which is a billion times the energy limit of current
terrestrial accelerators.  The energetic requirements at the source may
be even more stringent if the distance traveled by the UHE
primaries from source to Earth is larger than typical interaction
lengths. As can be seen from Figure 1 of
\cite{watson99} (from \cite{cronin92}), if $3 \times 10^{20}$ eV is
taken as a typical energy for protons travelling in straight lines,
accelerators located further than 30 Mpc need to reach above 1 ZeV while
those located further than 60 Mpc require over 10 ZeV energies. 
Depending on the strength and structure of the magnetic field along
the primary's path, the distance traveled can be  significantly larger
than the distance to the source. As magnetic fields above $\sim
10^{-8}$ G may thread extragalactic space \cite{RKB98,BBO99,FP99}, protons
travel in curved paths and sources need  to be either more energetic or
located closer to Earth \cite{WME96,BO99,SLB99}.

There are great difficulties with finding plausible accelerators for
such extremely energetic particles \cite{bland99}. As discussed in \S3,
even the most powerful astrophysical objects such as radio galaxies and
active galactic nuclei can barely accelerate charged particles to energies
as high as  $10^{20}$ eV.  If the origin of these events date back to the
early universe, then the energy is not as challenging since typical
symmetry breaking scales that give rise to early universe relics can be
well above the ZeV scale (\S5).

{\it 2.2 Flux}  

 At 10$^{20}$ eV, the observed flux of UHECRs is about $\sim$ 1
event/km$^2$/century which has strongly limited our ability to gather
more than 20 events after decades of observations
\cite{watson99}.  Although challenging to observers, the flux is not 
particularly constraining in terms of general requirements on
astrophysical sources. In fact, this flux equals the flux of
gamma-rays in  {\it one} gamma-ray burst that may have taken place in a
50 Mpc radius volume around us \cite{W95,V95}. In terms of an average
energy density, UHECRs correspond to $\sim 10^{-21}$ erg/cm$^3$, about
8 orders of magnitude less than the cosmic background radiation. 

Although less constraining to astrophysical accelerators, flux
requirements are very challenging for top-down scenarios. The dynamics
of topological defect generation and evolution generally selects the
present horizon scale as the typical distance between defects which
implies a very low flux. Some scenarios such as monopolia,  cosmic
necklaces, and vortons  have
additional scales and may avoid this problem.
The possibility of a long lived relic particle that cluster as dark
matter can also more easily meet the flux requirements than general
top-down models (\S5).

{\it 2.3 Spectrum} 

The energy spectrum of cosmic rays below the expected GZK cutoff (i.e.,
between $\sim 10^8\,$eV and $\la  10^{19}\,$eV) is well established  to
have a steep  energy dependence:  $N(E)  \propto E^{-\gamma}$, with
$\gamma\approx 2.7$ up to the ``knee'' at  $E\simeq 10^{15}$ eV and
$\gamma\approx 3.1$, for $10^{15}$ eV $\la  E \la 10^{19}$ eV.  Cosmic
rays of energy below the knee are widely accepted to originate in shocks
associated with galactic supernova remnants (see, e.g., \cite{A94}),
but this mechanism has difficulties producing particles of higher energies
\cite{NMA95}. Larger shocks, such as those associated with galactic
winds, could reach energies close to the knee \cite{JM87} and
supernova explosions into stellar winds may explain cosmic rays beyond
the knee \cite{SBG93}. Although the source of cosmic rays above the
knee is not clear, the steepening of the spectrum argues for a similar
origin with an increase in losses or decrease in  confinement time
above the knee. However, the events with energy above
$10^{19.5}\,$eV  show a much flatter spectrum with $1 \la  \gamma
\la 2$. The drastic change in slope suggests the emergence of a {\it
new component} of cosmic rays  at  ultra-high energies. This new
component is  generally thought to be  extragalactic \cite{A94,bird},
although, depending on its composition, it may also originate in the
Galaxy \cite{ZPPR98,OEB99}, in an  extended halo
\cite{V95}, or in the dark matter halo \cite{BKV97}.  Galactic and halo
origins for UHECRs ease the difficulties with the lack of a GZK cutoff but
represent an even greater challenge to acceleration mechanisms. 

{\it 2.4 Propagation - Losses and Magnetic Fields}

In order to contrast plausible candidates for UHECR sources with the
observed spectrum and arrival direction distribution,  the propagation
from source to Earth needs to be taken into account. Propagation
studies involve both the study of losses along the primaries' path as
well as the structure and magnitude of cosmic magnetic fields that
determine the trajectories of charged primaries and influence the
development of the electromagnetic cascade (see, e.g.,
\cite{LOS95PJ96}).

For primary protons the main loss processes are pair production 
\cite{Blu70} and photopion production off the CMB that gives rise to
the GZK cutoff \cite{GZK66}. For straight line propagation, loss
processes limit sources of 10$^{20}$ eV to be within $\sim$ 50
Mpc from us and a clear cutoff should be present at
$\sim 7 \times 10^{19}$ eV. Even with the small number of accumulated
events at the highest energies, the AGASA spectrum seems incompatible 
with a  GZK cutoff for a homogeneous extragalactic source distribution
\cite{takeda98}.  The shape of the cutoff can be modified if the
distribution of sources is not homogeneous \cite{BG88,MT99a} and if 
the particle trajectories are not rectilinear (e.g., the case of
sizeable intergalactic magnetic fields) 
\cite{WME96,S97,MT97,SLO97,BO99,SLB99}.  In fact, if the observed
distribution of galaxies in the local universe is used to simulate the
range of possible cutoff shapes, the AGASA spectrum is still 
consistent with sources distributed with the luminous matter given the
poor statistics
\cite{MT99a}.  The need for a new component should become apparent
with the increased statistics of future observatories \cite{watson99}. 

Charged particles of energies up to $10^{20}\,$eV can be deflected
significantly in cosmic magnetic fields. In a constant magnetic field
of strength $B= B_6 \mu$G, particles of energy $E= E_{20}10^{20}{\rm
eV}$ and charge $Ze$ have Larmor radii of
$r_L \simeq 110$ kpc $(E_{20}/B_6 Z)$.  If the UHECR primaries are
protons, only large scale intergalactic magnetic fields affect their
propagation significantly 
\cite{WME96,S97,MT97,SLO97,BO99,SLB99} unless the Galactic halo has 
extended fields \cite{S97}. For higher $Z$, the Galactic magnetic
field can strongly affect the trajectories of primaries
\cite{ZPPR98,HMR99}.

 Whereas Galactic magnetic fields are reasonably
well studied, extragalactic fields are still very
ill understood \cite{KronVallee}. Faraday rotation measures indicate large
magnitude fields   ($\sim
\mu$G) in the central regions  of clusters of
galaxies. In regions between clusters, the presence of
magnetic fields is evidenced by synchrotron emission but the strength
and structure are yet to be determined. On the largest
scales, limits can be imposed by the observed isotropy of the CMB
and by a statistical interpretation of Faraday rotation measures of light
from distant quasars. The isotropy of the CMB can 
constrain the present horizon scale fields
$B_{H_0^{-1}} \la  3 \times 10^{-9}$ G  \cite{BFS97}. Although the
distribution of Faraday rotation measures have large non-gaussian tails,
a reasonable limit can be derived using the  median of the  distribution
in an inhomogeneous universe: for fields assumed to be constant on the
present horizon scale, $B_{H_0^{-1}} \la  10^{-9}$ G; for fields with 50
Mpc coherence length,  $B_{50 {\rm Mpc}}
\la 6 \times 10^{-9}$ G; while for  1 Mpc coherence length, $B_{Mpc}
\la  10^{-8}$ G \cite{BBO99}. These limits apply to a $\Omega_b h^2 =
0.02$ universe and use quasars up to redshift $z=2.5$.  Local
structures can have fields above these upper limits as long as they are
not common along random lines of site between
$z$ = 0 and 2.5 \cite{RKB98,BBO99,FP99}.

Of particular interest is the field in the local 10 to 20 Mpc volume
around us.  If the Local Supercluster has fields of about
$10^{-8}$ G or larger, the propagation of ultra high energy protons
becomes diffusive and the spectrum and angular distribution at the
highest energies are significantly modified
\cite{WW79GWW80BGD89,RKB98,BO99}. As shown in Figure 1 (from
\cite{BO99}), a source with spectral index 
$\gamma\ga 2$ that can reach $E_{max} \ga 10^{20}$ eV is constrained by
the overproduction of lower energy events around 1 to 10 EeV (EeV $\equiv
10^{18}$ eV).
Furthermore, the structure and magnitude of magnetic fields in the
Galactic halo
\cite{S97,HMR99} or in a possible Galactic wind  can also affect
the observed UHECRs. In particular, if
our Galaxy has a strong magnetized wind, what appears to be an isotropic
distribution  in arrival directions may have
originated on a small region of the sky such as the Virgo cluster
\cite{ABMS99}. In the future, as sources of
UHECRs are identified,  large scale magnetic fields will be
better constrained \cite{LSOS97}.

\begin{figure}
\vspace{80mm}
\caption{Flux vs. Energy with $E_{max} = 10^{21}$ eV at source. Choices
of source distance r(Mpc), spectral index $\gamma$,  proton
luminosity $L_p$(erg/s), and LSC field B($\mu$G) are: solid line (13
Mpc,  2.1, $2.2 \times 10^{43}$ erg/s, 0.05 $\mu$G); dotted line (10
Mpc,  2.1, $10^{43}$ erg/s, 0.1 $\mu$G); dashed line (10 Mpc, 2.4, 3.2
$\times 10^{43}$ erg/s, 0.1 $\mu$G);  and dashed-dotted line (17 Mpc, 
2.1, $3.3 \times 10^{43}$ erg/s, 0.05 $\mu$G). Data points from  
\cite{Haya94,bird}.}
\end{figure}

If cosmic rays are heavier nuclei, the attenuation length is shorter 
than that for protons due to photodisintegration on the infrared
background \cite{PSB76SS99}. However, UHE nuclei may be of Galactic
origin. For large enough charge, the trajectories of UHE nuclei
are significantly affected by the Galactic magnetic field
\cite{HMR99} such that a Galactic origin can appear isotropic
\cite{ZPPR98}. The magnetically induced distortion of the flux
map   of UHE events can give rise to some higher flux regions where
caustics form and some much lower flux regions (blind spots) even for
an originally isotropic distribution of sources
\cite{HMR99}. Such propagation effects are one of the reasons why
full-sky coverage is necessary for resolving the UHECR puzzle.

The trajectories of neutral primaries are not affected
by magnetic fields. If associated
with luminous systems, sources of UHE neutral primaries should point back
to their nearby sources. The lack of counterpart identifications suggests
that if the primaries are neutral, their origin involves physics beyond
the standard model (\S4 \& \S5).

{\it 2.5 Cosmography} 

The distribution of arrival directions of UHECRs  can in principle hold
the key to solving the UHECR puzzle. Within a 50 Mpc radius volume
around us, the most well-known luminous structures are the  Galactic
plane, the Local Group and the large-scale galaxy distribution with a
relative overdensity around the Local
Supercluster. The Galactic halo is another noteworthy structure that
is expected to be a spheroidal overdensity of dark matter centered at
the Galactic disk while the dark matter distribution on larger scales
correlates with the luminous matter distribution.  For
the few highest energy events, there is presently no strong evidence
of correlations between the events' arrival direction and any of the 
known nearby luminous structures: the distribution is consistent with
isotropy
\cite{takeda99,SH99}.   For slightly lower
energies, some correlations may have been detected. For events around
40 EeV, a positive correlation with the Supergalactic plane is found but
only at the 1 $\sigma$ level \cite{Uchi99}. For even lower energies, a
more significant correlation was recently announced by  AGASA: the arrival
direction distribution of
 EeV events shows a correlation with the Galactic
center and the nearby Galactic spiral arms \cite{Haya99}. If
confirmed, this correlation would be strong evidence for a Galactic
origin of EeV cosmic rays.

{\it 2.6 Composition}

An excellent discriminator between proposed models is the composition
of the primaries. In general, Galactic disk models have to invoke
heavier nuclei such as iron to be consistent with the isotropic
distribution, while extragalactic
astrophysical models tend to favor proton primaries.  Photon primaries
are more common among top-down scenarios although nucleons can reach
comparable fluxes for some models \cite{BS99}. Experimentally, the 
composition can be determined by the muon content of the shower in 
ground arrays  and the depth of
shower maximum in fluorescence detectors \cite{watson99}. 
Unfortunately, the muon content analysis is not very  effective
at the highest energies. Data from the largest air shower array, AGASA,
disfavor photon primaries and indicate a fixed composition across the EeV
to 100 EeV range but does not distinguish  nuclei from proton primaries
\cite{Y99}. The shower development of the highest energy event ever
detected, the 320 EeV Fly's Eye event,  is consistent with either proton
or iron   \cite{bird} and also disfavors a photon primary
\cite{HVSV95}. This event constrains
hypothetical hadronic primaries to have masses below
$\sim $ 50 GeV  \cite{AFK98}.  Since fluctuations in shower
development are usually large, strong composition constraints await
larger statistics of future experiments.  
      
{\it 2.7 Clusters of Events}

A final challenge for models of UHECRs is the possible small scale
clustering of  arrival directions
\cite{Haya96,takeda99,Uchi99}. AGASA reported that
 their 47 events above 40 EeV show   three double coincidences
(doublets) and one triple coincidence (triplet) in arrival
directions,  a $\la 1$\% chance probability
\cite{takeda99}. Adding to the AGASA data that of   Haverah Park,
Volcano Ranch, and Yakutsk,  the 51 events above 50 EeV show  one
doublet and two triplets \cite{Uchi99}. Although these could be due to
a statistical fluctuation since the chance probability for the
combined set is $\sim 10 \%$
\cite{Uchi99}, they may indicate the position of the sources. (When
limited to $\pm 10^o$ around the Supergalactic plane the chance
probability decreases to $\sim 1\%$.) If these clusters
indicate the position of sources,  the arrival times and energies of
some of the events are inconsistent with a burst and require long lived
sources. Furthermore, if the clustering is confirmed by larger data
sets and their distribution correlates with some known matter
distributions in the nearby universe, the composition of the primaries
\cite{cronin96} as well as the magnitude of extragalactic magnetic
fields would be strongly constrained \cite{LSOS97,SLO97}. Alternative
explanations for the clustering involve either the effect of caustics
in the propagation due to magnetic fields \cite{HMR99} or the
clustering of dark matter in the halo of the Galaxy.

\section{Facing the Challenge with Zevatrons}

The challenge put forth by these observations has generated two different
approaches to reaching a solution: a `bottom-up' and a `top-down'. A
bottom-up approach involves looking for {\it Zevatrons}
\cite{bland99}, possible acceleration sites in known astrophysical
objects that can reach ZeV energies,  while a top-down
approach involves the decay of very high mass relics from the early
universe and physics beyond the standard model of particle physics.
Bottom-up models are discussed first and top-down models in the next
section.

\begin{figure}
\vspace{80mm}
\caption{$B$ vs. $L$, for $E_{max} =  10^{20}$ eV, $Z=1$
(dashed line) and $Z=26$  (solid line).}
\end{figure}

Acceleration of UHECRs in astrophysical plasmas occurs when large-scale
macroscopic motion, such as shocks and turbulent flows, is transferred
to individual particles. The maximum energy of accelerated  particles,
$E_{\rm max}$, can be estimated by requiring that the gyroradius of the
particle be contained in the acceleration region.  Therefore, for a
given strength, $B$, and coherence
length, $L$, of the magnetic field embedded in an astrophysical plasma,
$E_{\rm max} = Ze \, B \, L$, where  $Ze$ is the charge of the particle.
The ``Hillas plot''  \cite{hillas84} in Figure 2  shows that, for
$E_{max} \ga 10^{20}$ eV and $Z \sim 1$, the only known astrophysical
sources with reasonable  $B L $ products   are neutron stars ($B
\sim 10^{13}$ G, $L \sim 10$ km),  active galactic nuclei (AGNs) ($B
\sim 10^{4}$ G, $L \sim 10$ AU), radio lobes of AGNs ($B \sim 0.1\mu$G,
$L \sim 10$ kpc), and clusters of galaxies ($B \sim \mu$G, $L \sim
100$ kpc).

In general, when these sites are considered more carefully, one finds
great difficulties due to either energy losses in the acceleration
region or the great distances of known sources from our
Galaxy \cite{SSB94}. In many of these objects shock acceleration is
invoked as the primary acceleration mechanism. Although effective in
the acceleration of lower energy cosmic rays, shock acceleration is
unable to reach ZeV energies for most plausible acceleration sites
\cite{NMA95} with the possible exception of shocks in
radio lobes. Unipolar inductors are often
invoked as plausible alternative to shocks 
\cite{BBDGP90,bland99}.

\smallskip
{\it 3.1 Cluster Shocks}

Moving from right to left on Figure 2, cluster shocks are reasonable
sites to consider for UHECR acceleration, since $E_{max}$ particles
can be contained by cluster fields. However, the propagation of these
high energy particles inside the cluster medium is such that they do
not escape without significant energy losses.  In fact, efficient
losses occur on the scales of clusters of galaxies for the same reason
that a GZK cutoff is expected, namely, the photopion production off
the CMB. Losses limit UHECRs in cluster shocks  to reach at most
$\sim$ 10 EeV \cite{KRJ96KRB97}.

\smallskip
{\it 3.2 AGN - Jets and Radio Lobes}

Extremely powerful radio galaxies are likely astrophysical
UHECR accelerators \cite{hillas84,BS87} (for a
recent review see \cite{B97}).  Jets from the central black-hole of
the active galaxy end at a termination shock where the interaction of
the jet with the intergalactic medium forms radio lobes and  `hot
spots'. Of special interest are the most powerful AGNs such as
Fanaroff-Riley class II objects \cite{FR74}. Particles accelerated in
hot spots of FR-II sources via first-order Fermi acceleration may reach
energies well above an EeV  and may explain  the spectrum up to the GZK
cutoff \cite{RB93}. A nearby specially powerful source
 may be able to reach energies  past the
cutoff \cite{RB93}.  Alternatively, the crossing of the
tangential discontinuity between the relativistic jet and the
surrounding medium may also be able to make protons reach the necessary
energies \cite{Ost99}.  The spectrum of UHECR primaries formed by the
latter proposal is flatter than the Fermi acceleration at the hot spots
scenario.  Improved statistics of events past the GZK cutoff by future
experiments should better determine the  spectral index, and therefore, 
discriminate between plausible sites for UHECR
acceleration in radio sources. 

Both  hot spots and tangential jet
discontinuity models avoid the efficient loss processes faced by
acceleration models in AGN central regions (\S 3.3).
However,  the location of possible sources is problematic
for both types of mechanisms. Extremely powerful AGNs
with radio lobes and hot spots are rare and far apart. The closest known
object is M87 in the Virgo cluster ($\sim$ 18 Mpc away)  and could be
a main source of UHECRs.  Although a single nearby source may be able
to fit the spectrum for a given strength and structure of the
intergalactic magnetic field
\cite{BO99}, it is unlikely to match the observed arrival direction
distribution. After M87, the next known nearby source is NGC315 which
is already too far  at a distance of
$\sim $ 80 Mpc. 

A recent proposal gets around
this challenge  by invoking a  Galactic wind with a  strongly magnetized
azimuthal component \cite{ABMS99}. Such a wind can significantly alter
the paths of UHECRs such that all the observed arrival directions of
events above 10$^{20}$ eV trace back to the Virgo cluster close to  M87
 \cite{ABMS99}. If our Galaxy has a wind  with
the required characteristics to allow for this magnetic focusing is
yet to be determined. Future observations of UHECRs from the Southern
Hemisphere  (e.g., the Southern Auger Site
\cite{cronin92}) will provide data on
previously unobserved parts of the sky  and help   distinguish plausible
proposals for the effect of local magnetic fields on arrival
directions. Once again full sky coverage is a key
discriminator of such proposals.

 \smallskip
{\it 3.3 AGN - Central Regions}

The powerful engines that give rise to the observed jets and radio
lobes are located in the central regions of active galaxies and are
powered by the accretion of matter onto supermassive black holes. It
is reasonable to consider the central engines themselves as the likely
accelerators
\cite{hillas84,T86,BBDGP90}. In principle, the nuclei of  generic
active galaxies (not only the ones with hot spots) can accelerate
particles via a unipolar inductor 
\cite{T86} not unlike the one operating in pulsars
\cite{GJ69}. In the case of AGNs,   the magnetic field  is provided by
the infalling matter and the spinning black hole horizon provides the
imperfect conductor for the unipolar induction. Close to the horizon
of  a black hole ($R\simeq GM/c^2$) with a mass M
$= 10^{9} M_9 \ {\rm M}_{\odot}$, the electromotive force is
\cite{BZ77,T86}:
$emf \propto cBR \approx 4.4 \times 10^{20} B_4 M_9 {\rm Volts} $ 
for  a magnetic field $B=10^4 B_4$ G. It is reasonable to expect that
such fields are reached in some  nearby AGNs.   
In addition,  the arrival direction of events above $5 \times 10^{19}$
eV correlate qualitatively well with active galaxies within 100 Mpc
\cite{Cro99}. Although it is not clear how statistically significant
the correlation is, the clustering of UHECR events in the same regions of
the sky where clusters of AGNs reside is certainly tantalizing.

The problem with AGNs as UHECR sources is two-fold: first, UHE particles
face  debilitating losses in the acceleration region due to the intense
radiation field present in AGNs,  and second, the
spatial distribution of objects should give rise to a  GZK cutoff of the
observed spectrum. In the central  regions of AGNs, loss processes are
expected to downgrade particle energies well below the maximum
achievable energy. This limitation has led to the proposal that  quasar
remnants, supermassive black holes in centers of inactive galaxies,  are
more effective UHECR accelerators
\cite{BG99}. In this case, losses are not as significant. In addition,
the problem with the rarity of very luminous radio sources (\S 3.2) is
also avoided since  any galaxy with a supermassive quiescent black hole
could host a UHECR accelerator.

Quasar remnants are manifestly underluminous such that
 losses in the acceleration region are kept at
a reasonably low level \cite{BG99}. Although presently underluminous,  
the underlying supermassime black holes  are likely to be sufficiently 
spun-up for individual particles to be accelerated. 
An incomplete sample of 32  massive dark objects
(MDOs) in the nearby universe
(of which 8 are within 50 Mpc) \cite{Ma98} finds about 14 MDOs 
which  could have fields strong enough for an $emf \ga 10^{20}$
Volts \cite{BG99}.  From the number density and accretion evolution of
quasars, more than 40 quasar remnants are expected to have
$\ga 4 \times 10^{8}$ M$_{\odot}$ within a 50
Mpc volume while more than a dozen would have $\ga 10^{9}$
M$_{\odot}$ \cite{CT92}.

The second difficulty with AGNs mentioned above, namely the spatial
distribution and the GZK cutoff induced by the more distance galaxies, is
not avoided by the  quasar remnants proposal unless the spectrum is
fairly hard. However, it is still within the errors of the current
UHECR spectrum the possibility that a GZK cutoff is presently hidden
due to the effect of the local clustering of galaxies
\cite{MT99a}. This ambiguity should be lifted and a GZK cutoff made
apparent by future experiments.

 \smallskip
{\it 3.4  Neutron Stars}

From Figure 2, the last astrophysical objects capable of accelerating
UHECRs are neutron stars (see, e.g., \cite{hillas84,BBDGP90}).
With the recent identification of ``magnetars'' \cite{TD95} (neutron
stars with fields of $\ga 10^{14}$ G) as the sources of soft gamma
ray repeaters  \cite{Kou98}, neutron stars have strong enough fields
to  reach past the required $E_{max}$ as in Figure 2.  Acceleration
processes inside the neutron star light cylinder are bound to fail
much like the AGN central region case:  ambient magnetic and radiation
fields induce significant losses 
\cite{VMO97}. However, the plasma that expands beyond the light
cylinder is freer from the main loss processes and may be accelerated
to ultra high energies.
One possible solution to the UHECR puzzle is the proposal that
the early evolution of neutron stars may be responsible for the flux
of cosmic rays beyond the GZK cutoff 
\cite{Be92,OEB99,BEO99}. In this case, UHECRs originate mostly in the
Galaxy and the arrival directions require that the primaries have large
$Z$ (i.e., primaries are heavier nuclei).

Newly formed, rapidly rotating neutron stars may accelerate iron
nuclei  to UHEs  through relativistic MHD winds beyond 
their light cylinders \cite{OEB99,BEO99}.  The nature  of the
relativistic wind is not yet clear, but observations of the Crab
Nebula indicate that most of the rotational energy emitted by  the
pulsar is converted into the flow kinetic energy of the particles in
the wind (see, e.g., \cite{BL98}). Recent observations of the Crab Nebula
by the Chandra satellite indicate both a complex disk and jet structure
that is probably associated with the magnetic wind as well as the presence
of iron in the expanding shell. Understanding the structure
of observable pulsar winds such as the Crab nebula will help determine if
during their first years pulsars were efficient Zevatrons.

If most of the magnetic energy in
the wind zone is converted into  particle kinetic energy  and  the rest
mass density of the wind is not dominated by electron-positron pairs,  
 particles  in the wind can reach a maximum energy of
$E_{max} \simeq 8 \times 10^{20}  \, Z_{26} B_{13} \Omega_{3k}^2 \,
{\rm eV}, $ for iron nuclei  ($Z_{26} \equiv Z/26 = 1)$, neutron star
surface fields $B = 10^{13} B_{13} $ G,  and initial rotation frequency
$\Omega = 3000 \Omega_{3k}$ s$^{-1}$. In the rest frame of the wind,
the plasma is relatively cold while in the star's  rest frame the
plasma moves with Lorentz factors $\gamma \sim 10^9 - 10^{10}$. 
 
Iron nuclei can
escape the remnant of the supernova without suffering significant
spallation about a year after the explosion.  As the
ejected envelope of the pre-supernova star expands,  the young neutron
star spins down and $E_{max}$ decreases. Thus, a requirement for
relativistic winds to  supply  UHECRs is that the column density of
the envelope becomes transparent to UHECR iron  before the spin rate
of the neutron star decreases significantly. The allowed parameter
space for this model is shown in Figure 3. Magnetars with the
largest surface fields spin down too quickly for iron nuclei to escape
unless the remnant is asymmetric with  lower density ``holes.'' The
spectrum of UHECRs accelerated by young neutron star winds is
determined by the evolution of the  rotational frequency which gives
$\gamma \simeq 1$, at the hard end of the allowed $\gamma$ range
(\S 2.3). 

\begin{figure}
\vspace{80mm}
\caption{Allowed regions of $\Omega$ vs. $B$ for $E_{cr}=10^{20}$ eV
(solid line) and $ 3\times 10^{20}$ eV (dashed lines) with envelope
masses $M_{env}=50 M_{\odot}$ and $5 M_{\odot}$.  Horizontal line
indicates the minimum period for neutron stars  $\sim 0.3$ ms.}
\end{figure}

Depending on the structure of   Galactic
magnetic fields, the trajectories of iron nuclei from Galactic
neutron stars may be consistent with the observed arrival directions of
the highest energy events \cite{ZPPR98}. Moreover,  if  cosmic rays 
of a few times $10^{18}$ eV are protons of Galactic origin, the
isotropic distribution observed at these energies is indicative of the
diffusive effect of the Galactic magnetic fields on iron at
$\sim 10^{20}$ eV. 

Another recent proposal involving neutron stars suggests that
relativistic  winds formed around neutron star binaries may generate
high energy cosmic rays in a single shot $\Gamma^2$ acceleration  
\cite{GA99}, where $\Gamma$ is the  bulk Lorentz factor. However, the
$\Gamma^2$ acceleration process is likely to be very inefficient 
which renders the proposal insufficient for explaining UHECRs
\cite{BeOs99}. 
 
In general, there is an added bonus to considering the existence of
Zevatrons in Galactic systems: one may find Pevatrons or Evatrons
instead. These may explain the  origin of cosmic rays from  the
knee at 10$^{15}$ eV up to the ``ankle'' at  10$^{18}$ eV that remain
largely  unidentified. 

 \smallskip
{\it 3.5 Gamma-Ray Bursts}

Before moving on to more exotic explanations for the origin of UHECRs,
one should consider astrophysical  phenomena that
may act as Zevatrons  not included in Figure
2.  In effect, transient high energy phenomena such as gamma-ray
bursts  (bursts of $\sim 0.1 - 1 $ MeV photons that last up to a few 
seconds) may accelerate protons to ultra-high energies 
\cite{W95,V95}. The systems that generate gamma-ray
bursts (GRBs) remain unknown
but evidence that GRBs are of cosmological origin and involve a
relativistic fireball has been mounting with the recent discovery of
X-ray, optical, and radio afterglows  \cite{C97G97F97} and the
subsequent identification of host galaxies and their redshifts.

Aside from both having unknown origins, GRBs and UHECRs have some
similarities that argue for a common origin. Like UHECRs, GRBs are
distributed isotropically in the sky  \cite{BATSE92},  and
the average rate of $\gamma$-ray energy emitted by GRBs is comparable
to the energy generation rate of UHECRs of energy $>10^{19}$ eV in a
redshift independent cosmological distribution of sources
\cite{W95}, both have  $ 
\approx 10^{44}{\rm erg\ Mpc}^{-3}{\rm yr}^{-1} .$ 

Although the systems that generate GRBs 
have not been identified, they are likely to involve a
relativistic fireball   (see, e.g., \cite{fireballs}).  Cosmological
fireballs may generate UHECRs through Fermi acceleration
by internal shocks
\cite{W95,V95}. In this model the
generation spectrum is estimated to be 
$dN/dE\propto E^{-2}$ which is  consistent with observations
provided the efficiency with which the wind
kinetic energy is converted to $\gamma$-rays is similar to the
efficiency with which it is converted to UHECRs \cite{W95}. 
 Acceleration to $>10^{20}$ eV is possible provided
that  $\Gamma$ of  the fireball shocks are large enough
and that the magnetic field is close to equipartition.

There are a few  problems with the GRB--UHECR common origin
proposal. First, events past the GZK cutoff require that only GRBs
from $\la 50$ Mpc contribute. However, only {\it one} burst is
expected to have occurred within this region over a period of 100 yr.  
Therefore, a very large dispersion of   $\ga$ 100  yr  in
the arrival time of protons  produced in a single burst is a necessary
condition. The deflection  by random
magnetic fields combined with the energy spread of the particles
is usually invoked to reach the required dispersion
\cite{W95,WME96}. If the dispersion in time is achieved, the energy
spectrum for the nearby source(s) is expected to be very narrowly 
peaked  $\Delta E/E\sim1$ \cite{W95,WME96,LSOS97}.
Second, the fireball shocks may not be able to reach the
required $\Gamma$ factors for UHECR shock acceleration \cite{GA99}.
Third, UHE protons are likely to loose most of their energy as they
expand adiabatically with the fireball \cite{RM98}. However, 
if acceleration happens by internal shocks in regions where the
expansion becomes self-similar, protons may escape without significant
losses \cite{W99}. Fourth, the  observed arrival times of different
energy events in some of the UHE clusters
  argues for long lived sources not bursts (\S 2.7). These clusters can
still be due to fluctuations but should become clear in future experiments
\cite{SLO97}. Finally, the present flux of UHE protons from GRBs is
reduced to
$\la 10^{42}{\rm erg\ Mpc}^{-3}{\rm yr}^{-1}$, if a redshift dependent
source distribution that fits the GRB data  is
considered \cite{Ste99} (see also \cite{FP99,D99}).

\section{Hybrid Models}

The UHECR puzzle has inspired proposals that use Zevatrons to
generate UHE particles other than protons, nuclei, and photons.
These use physics beyond the standard model in a bottom-up approach,
thus, named hybrid models.

The most economical among such proposals involves a familiar
extension of the standard model, namely, neutrino masses. The most 
common solution to  the atmospheric  or the solar neutrino
problems entails  neutrino oscillations, and hence,  neutrino
masses (see, e.g., \cite{B89}).
Recently, the announcement by SuperKamiokande on atmospheric neutrinos 
has strengthened the evidence for neutrino oscillations and the
possibility that neutrinos have a small mass \cite{SK99}. If
some flavor of neutrinos have masses $\sim 1$ eV, the relic
neutrino background  will cluster in halos of galaxies and clusters of
galaxies. High energy neutrinos ($\sim 10^{21}$ eV) accelerated in
 Zevatrons  can annihilate on the neutrino background and
form UHECRs through the hadronic Z-boson decay \cite{We97FMS97}. 

This proposal is
aimed at generating UHECRs nearby (in the Galactic halo and Local Group
halos) while using Zevatrons that can be much further than the GZK
limited volume, since neutrinos do not suffer the GZK losses. It is not
clear if the goal is actually achieved since the production in the
uniform non-clustered neutrino background may be comparable to 
the local production depending on the neutrino masses
\cite{Wa99}. In addition, the Zevatron needed to accelerate protons above
ZeVs that can produce ZeV neutrinos as secondaries is quite spectacular and
presently unknown, requiring an energy generation in excess of
$\sim 10^{48} {\rm erg\ Mpc}^{-3}{\rm yr}^{-1}$ \cite{Wa99}.

Another suggestion is that the UHECR primary is a new 
particle. For instance,  a stable or very long
lived supersymmetric neutral hadron of a few GeV, named {\it
uhecron}, could explain the UHECR events and evade the present
laboratory bounds
\cite{Fa96CFK98}. (Note that the mass of a hypothetical hadronic
UHECR primary can be limited by the shower development of the Fly's
Eye highest energy event to be
 below $\la 50$ GeV \cite{AFK98}.)  Both the long lived new
particle and the neutrino Z-pole proposals involve neutral particles which
are usually harder to accelerate (they are created as secondaries of even
higher energy charged primariess) but
can traverse large distances without being affected by the cosmic magnetic
fields. Thus, a signature of such hybrid models for future experiments is
a clear correlation between the position of powerful Zevatrons in the sky
such as distant compact radio quasars and the arrival direction of  UHE
events
\cite{FB98}.

Topological defects have also been suggested as possible UHE primaries
\cite{P60}. Monopoles of masses between $\sim 10^{9} - 10^{10}$ GeV 
have relic densities below the Parker limit and can be easily
accelerated to ultra high energies by the Galactic magnetic field
\cite{KW96}.  The main challenges to this proposal are the observed 
shower development for the Fly's Eye event that seems to be
inconsistent with a monopole primary and the arrival directions not
showing a preference for the local Galactic magnetic field \cite{MN98}.

Another exotic primary that can use a Zevatron to reach ultra high
energies is the vorton. Vortons are small loops of superconducting
cosmic string stabilized by the angular momentum of charge carriers
\cite{DS89}. Vortons can be a component of the dark
matter in galactic halos and be accelerated in astrophysical
Zevatrons \cite{BP97}. Although not yet clearly demonstrated, the
shower development profile is also the likely liability of this model.

\section {Top-Down Models}

It is possible that none of the astrophysical scenarios are able to 
meet the challenge posed by the UHECR data as more
observations are accumulated. In that case,  the alternative
is to consider top-down models. For example, if the primaries are
not  iron, the distribution in the sky remains isotropic with better
statistics, and the spectrum does not show a GZK cutoff, UHECRs are
likely to be due to the decay of very massive relics from the early
universe.

 This possibility was the most attractive to my dear
colleague and friend, David N. Schramm, to whom this volume is
dedicated.  After learning with the work of Hill \cite{H83} that high
energy particles would be produced by the decay of supermassive Grand
Unified Theory (GUT) scale particles (named X-particles)
in monopole-antimonopole annihilation, Schramm joined Hill in proposing
that such processes would be observed as the highest energy cosmic
rays \cite{SH83HS83}.  Schramm  realized the
potential for explaining UHECRs with physics at very high energies well
beyond those presently available at terrestrial accelerators. One
winter in Aspen, CO, he remarked pointing to the ski lift `why walk up
if we can start at the top'. His
 enthusiasm for this problem only grew after his pioneering work
\cite{HS8597}. In the last
conference he attended, an OWL workshop at the University of Maryland
 \cite{Owl97}, he summarized the meeting by reminding us
that in this exciting field  the most conventional proposal involves
supermassive black holes and that the best fit models involve physics
at the GUT scale and beyond. In this field our imagination is the
limit (as well as the low number of observed events). 

The lack of a clear astrophysical solution for
the UHECR puzzle has encouraged a number of interesting proposals based
on physics beyond the standard model such as monopolia annihilation,
the decay of ordinary and superconducting cosmic strings,  cosmic
necklaces, vortons, and superheavy long-lived relic particles, to name
a few. Due to the lack of space and a number of recent thorough
reviews, only a brief summary of the general features of
these proposals will be given here. The interested reader is encouraged
to consult the following  reviews by long-time collaborators of David
Schramm \cite{berez99,BS99} and references therein.

The idea behind top-down models is that relics of the very
early universe, topological defects (TDs) or superheavy relic (SHR)
particles,  produced  after or at the end of inflation, can
decay today and generate UHECRs.  Defects, such as cosmic strings,
domain walls, and magnetic monopoles,  can be generated through the
Kibble mechanism
\cite{Ki76}  as symmetries are broken with the
expansion and cooling of the universe (see, e.g., \cite{TDs}). 
Topologically stable defects can survive to the  present and
decompose into their constituent fields  as they collapse, 
annihilate, or reach critical current in the case of superconducting
cosmic strings. The decay products, superheavy gauge and higgs bosons,
decay into jets of hadrons, mostly pions.  Pions in the jets
subsequently decay into $\gamma$-rays, electrons, and neutrinos. Only a
few percent of the hadrons are expected to be nucleons \cite{H83}.
Typical features of these scenarios are a predominant release of
$\gamma$-rays and neutrinos and a QCD
fragmentation spectrum which is  considerably harder than the case of
shock acceleration.

ZeV energies are not a challenge for top-down models since symmetry
breaking scales at the end of inflation typically are $\gg 10^{21}$
eV (typical X-particle masses vary between 
$\sim 10^{22} - 10^{25}$ eV) .  Fitting the observed flux
of UHECRs is the real challenge since the typical distances between TDs
is  the  Horizon scale,
$H_0^{-1} \simeq 3 h^{-1}$ Gpc. The low flux hurts proposals based on
ordinary  and superconducting cosmic strings 
\cite{berez99,BS99}. Monopoles usually suffer the opposite
problem, they would in general be too numerous. Inflation succeeds in
diluting the number density of monopoles \cite{G81} 
usually making them too rare for UHECR production. To reach the
observed UHECR flux, monopole models usually involve some degree of
fine tuning. If enough monopoles and antimonopoles survive from the
early universe,  they can form a bound state, named monopolium, 
that  decay  generating UHECRs  through   monopole-antimonopole
annihilation   \cite{H83,BS95}. The lifetime of monopolia may be too short
for this csenario to succeed unless they are connected by
strings \cite{PO99}.

Once two symmetry breaking scales are invoked, a combination of
horizon scales gives room to reasonable number densities. This can be
arranged for cosmic strings that end in monopoles making a monopole
string network or even more clearly for cosmic necklaces
\cite{BV97}. Cosmic necklaces are hybrid defects where each
monopole is connected to two strings resembling beads on a cosmic
string necklace. Necklace networks may evolve to configurations that
can fit the UHECR flux which is ultimately generated by the
annihilation of monopoles with antimonopoles trapped in the string
\cite{BV97,BBV98}. 

\begin{figure}
\vspace{80mm}
\caption{Proton and $\gamma$-ray fluxes from necklaces for 
$m_X=  10^{14}$ GeV (dashed lines),  $10^{15}$ GeV (dotted 
lines), and $10^{16}$ GeV (solid lines) normalized to
the  observed data.
$\gamma$-high   and  $\gamma$-low  correspond to two extreme cases 
of $\gamma$-ray absorption (see, \cite{BBV98}).}
\end{figure}

In addition to fitting the UHECR flux, topological defect
models are constrained by limits on the flux of  high energy photons
observed by EGRET (10 MeV to 100 GeV). The energy density of lower
energy cascade photons generated by UHE photons and electrons off the
CMB and radio background   is limited  to $\la 10^{-6}$ eV/cm$^3$. 
Figure 4  shows the predicted flux for necklace
models  given different radio backgrounds and different masses for the
X-particle (from \cite{BBV98}). As can be seen from the Figure, protons
dominate the flux at lower energies while photons tend to dominate at
higher energies depending on the radio background. If future data can
settle the composition of UHECRs from 0.01 to 1 ZeV, these models will
be well constrained.

Another interesting possibility is the recent proposal that UHECRs are
produced by the decay of unstable superheavy relics that live much
longer than the age of the universe \cite{BKV97,KR97}. 
SHRs may be produced at the end of inflation by non-thermal effects
such as a varying gravitational field, parametric resonances during
preheating,  instant preheating, or the decay of topological defects
(see, e.g., \cite{KT99}). 
SHRs have unusually long lifetimes insured by
discrete gauge symmetries  and a sufficiently
small percentage decays today producing UHECRs \cite{BKV97,CKR99KT99}.
As in the topological defects case, the decay of these relics also
generate jets of hadrons. 
These particles behave like cold
dark matter and could constitute a fair fraction of the halo of our
Galaxy. Therefore, their halo decay products would not be limited by
the GZK cutoff allowing for a large flux at UHEs.
 The flux of UHECRs predicted by  SHRs clustered in our halo
is plotted in Figure 5 (from \cite{BBV98}). It is clear that the
spectrum is not power law (unlike the case of shock acceleration) and
that photon fluxes dominate. 

\begin{figure}
\vspace{80mm}
\caption{SHRs  or monopolia decay fluxes
(for $m_X= 10^{14} ~GeV$):
nucleons from the halo ({\it protons}), $\gamma$-rays
from the halo ({\it gammas}) and extragalactic protons. Solid, dotted
and dashed curves correspond to different  model parameters
(see \cite{BBV98}).}
\end{figure}

From Figures 4 and 5 it is clear that future experiments should be
able to probe these hypotheses.  For instance, in the case of SHR
and monopolium decays, the arrival
direction distribution should be close to isotropic but show an
asymmetry due to the position of the Earth in the Galactic Halo
\cite{BBV98,DT98BSW99BM99}. Studying plausible halo models and the
expected asymmetry will help constrain halo distributions especially
when larger data sets are available from future experiments. High
energy gamma ray experiments such as GLAST will also help constrain
the SHR models due to the products of the electromagnetic cascade
\cite{B99}.

\section{Conclusion}

Next generation experiments  such as the High Resolution Fly's Eye
\cite{hires2} which recently started operating, the Pierre Auger
Project \cite{cronin92} which is now under construction,  the
proposed  Telescope Array \cite{Teshima92}, and
the OWL-Airwatch satellite \cite{St97} will 
significantly improve the data at the extremely-high end of the cosmic
ray spectrum \cite{watson99}. With these observatories a clear
determination of the spectrum and spatial distribution of UHECR
sources is within reach. 
The lack of a GZK cutoff should become apparent with Auger
\cite{MT99a} and most extragalactic Zevatrons may be ruled out.
 The observed spectrum will distinguish Zevatrons from
top-down models by testing power laws versus QCD fragmentation fits. 
The cosmography of sources should also become clear and able to
 discriminate  between plausible populations for UHECR sources. 
The  correlation of arrival directions  for events with
energies above
$10^{20}$ eV  with some known structure such as the Galaxy, the
Galactic halo, the Local Group or the Local Supercluster would be key
in differentiating between different models. For instance, a
correlation with the Galactic center and  disk should become apparent
at extremely high energies for the case of young neutron star winds
\cite{SH99}, while a correlation with the large scale galaxy
distribution should become clear for the case of quasar remnants.
If SHRs or monopolia are responsible for UHECR production, the arrival
directions should correlate with the dark matter distribution and show
the halo asymmetry. For these signatures to be tested, full sky
coverage is essential. Finally,  an excellent discriminator would be
an unambiguous composition determination  of the primaries. In
general, Galactic disk models  invoke iron nuclei to be consistent
with the isotropic distribution,  extragalactic Zevatrons
tend to favor proton primaries, while photon primaries are more common
for early universe relics. The hybrid detector of the Auger Project
should help determine the composition by measuring simultaneously
the depth of shower maximum and the muon content of the same
shower.

In addition to explaining the origin of UHECRs, GUT to Planck scale
physics can potentially be probed by the existence of UHECRs. For
instance, the breaking of Lorentz invariance can change the
threshold for photopion production significantly in such a way as to
be constrained by a clear observation of the GZK cutoff
\cite{ABGG00}.   There are great gains to be made if the data at the
highest energies is improved by a few orders of magnitude. The
prospect of testing extremely high energy physics as well as solving
the UHECR puzzle given all the presently proposed  models sends a
strong message that the challenge is back in the observational arena.
Fortunately, observers have accepted the challenge and  
are building and planning  experiments  large
enough to resolve  these open questions
\cite{watson99}.

\section*{Acknowledgment}
 
It has been a great pleasure to have known David N. Schramm and to be
able to contribute to this volume in his honor. Dave was a kind 
mentor and friend who is missed with {\it saudades.} Many thanks to
my ``ultra-high energy'' collaborators, P. Blasi and R. Epstein,
for the careful reading of the manuscript, many ongoing discussions,
and for providing most of the figures. I am also very grateful to I.
Albuquerque, V. Berezinsky, P. Biermann,  J. Cronin, G. Farrar, T. Gaisser,
M. Lemoine,  G. Sigl, T. Stanev, A. Watson, and T. Weiler for their
comments on the manuscript. This work was supported by NSF through grant
AST 94-20759  and DOE grant DE-FG0291  ER40606.

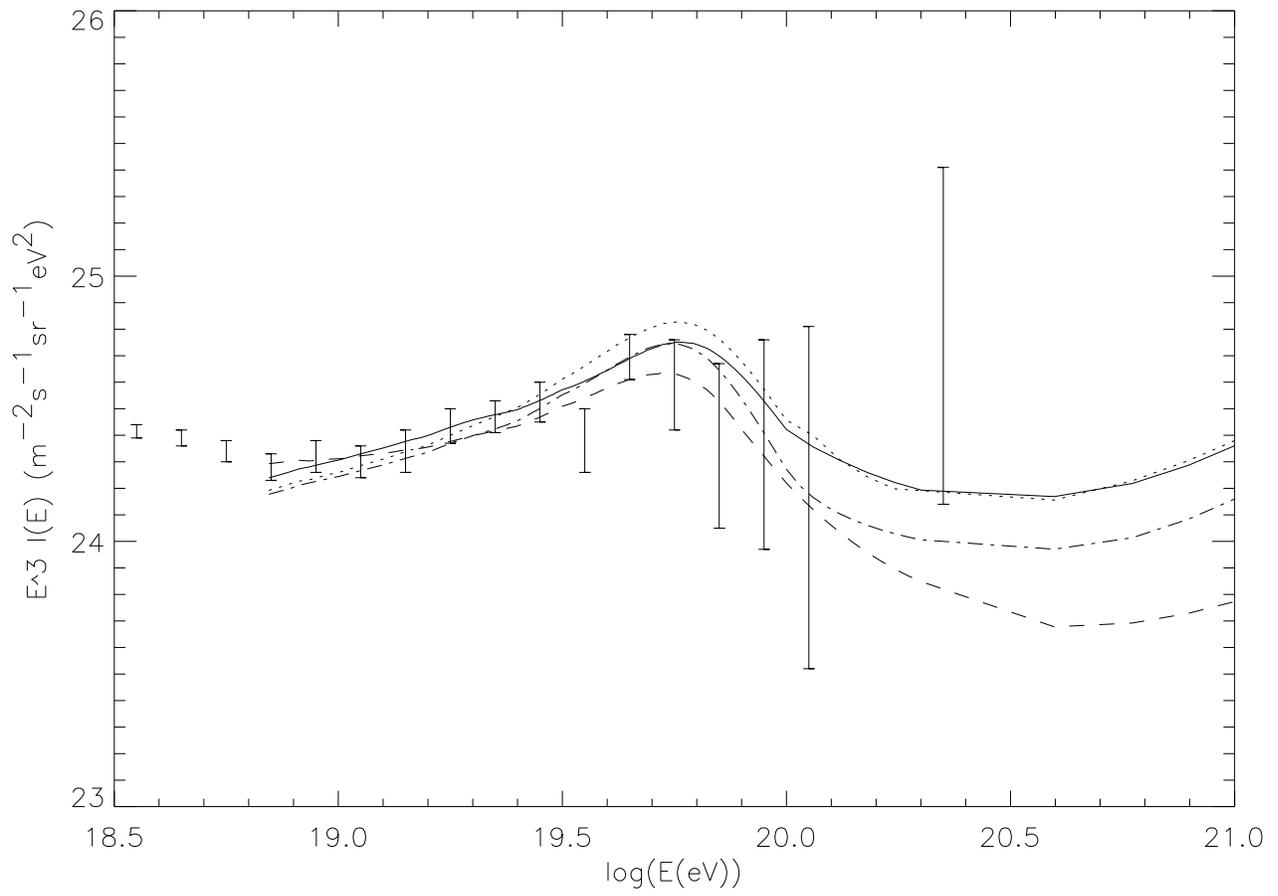
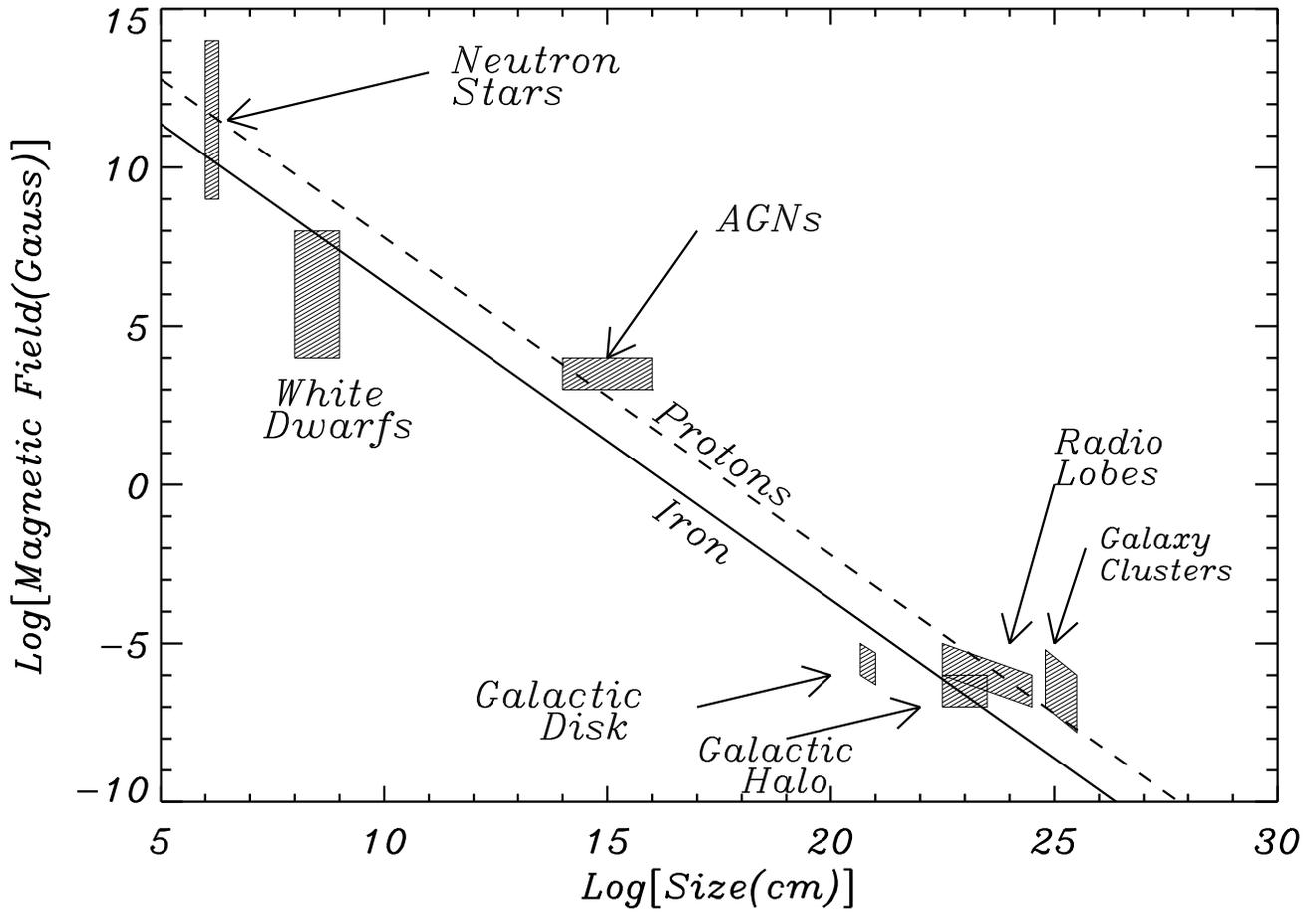
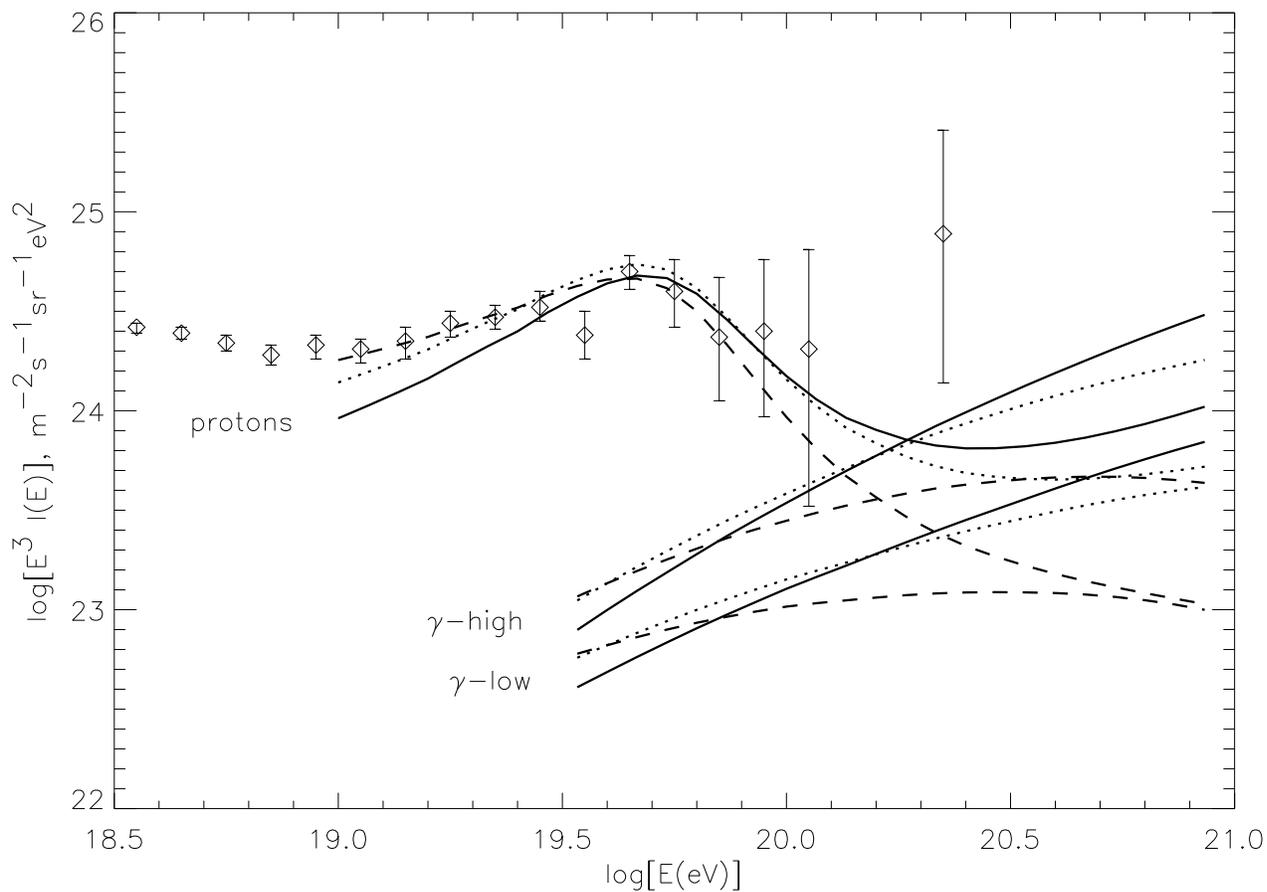
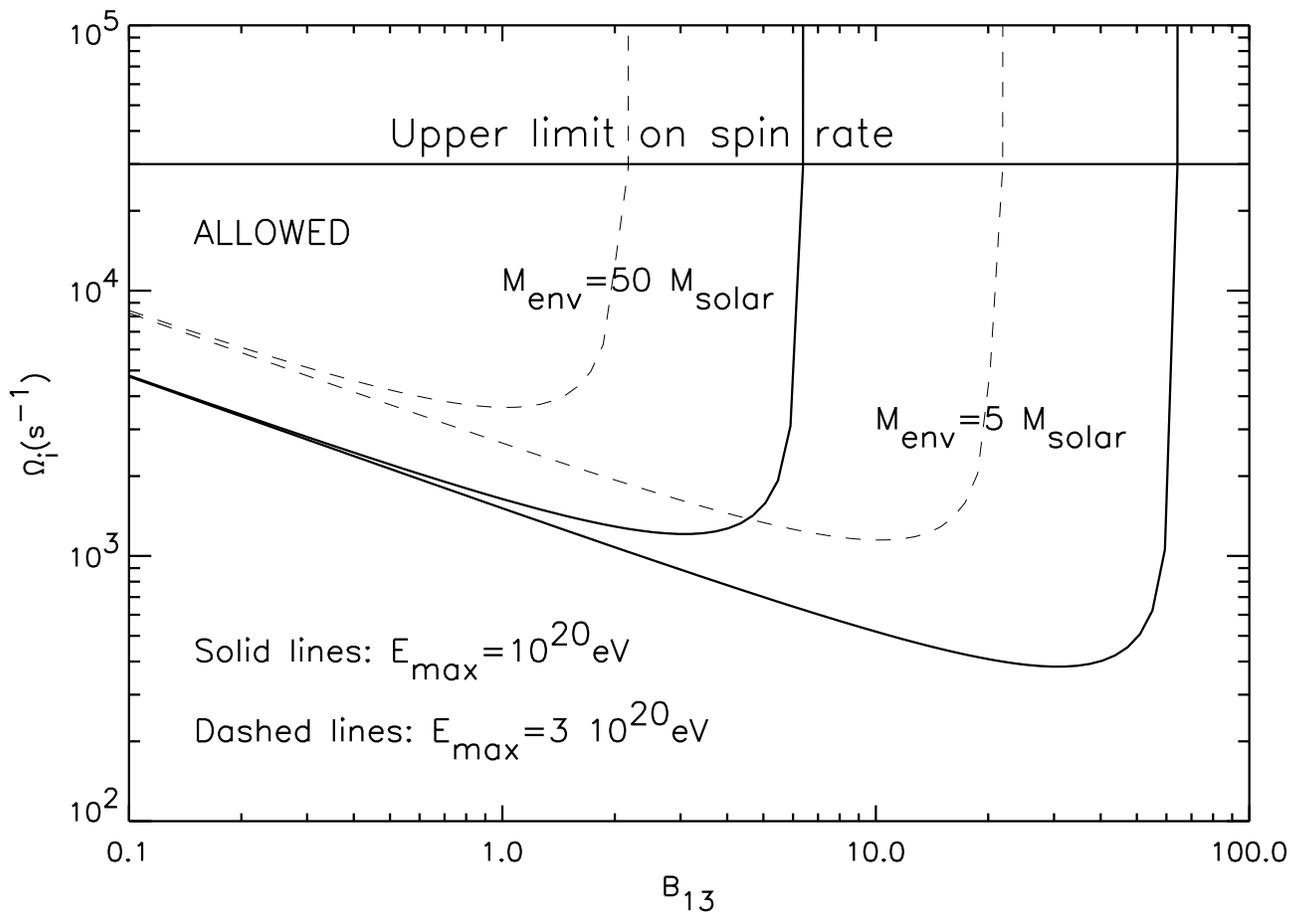
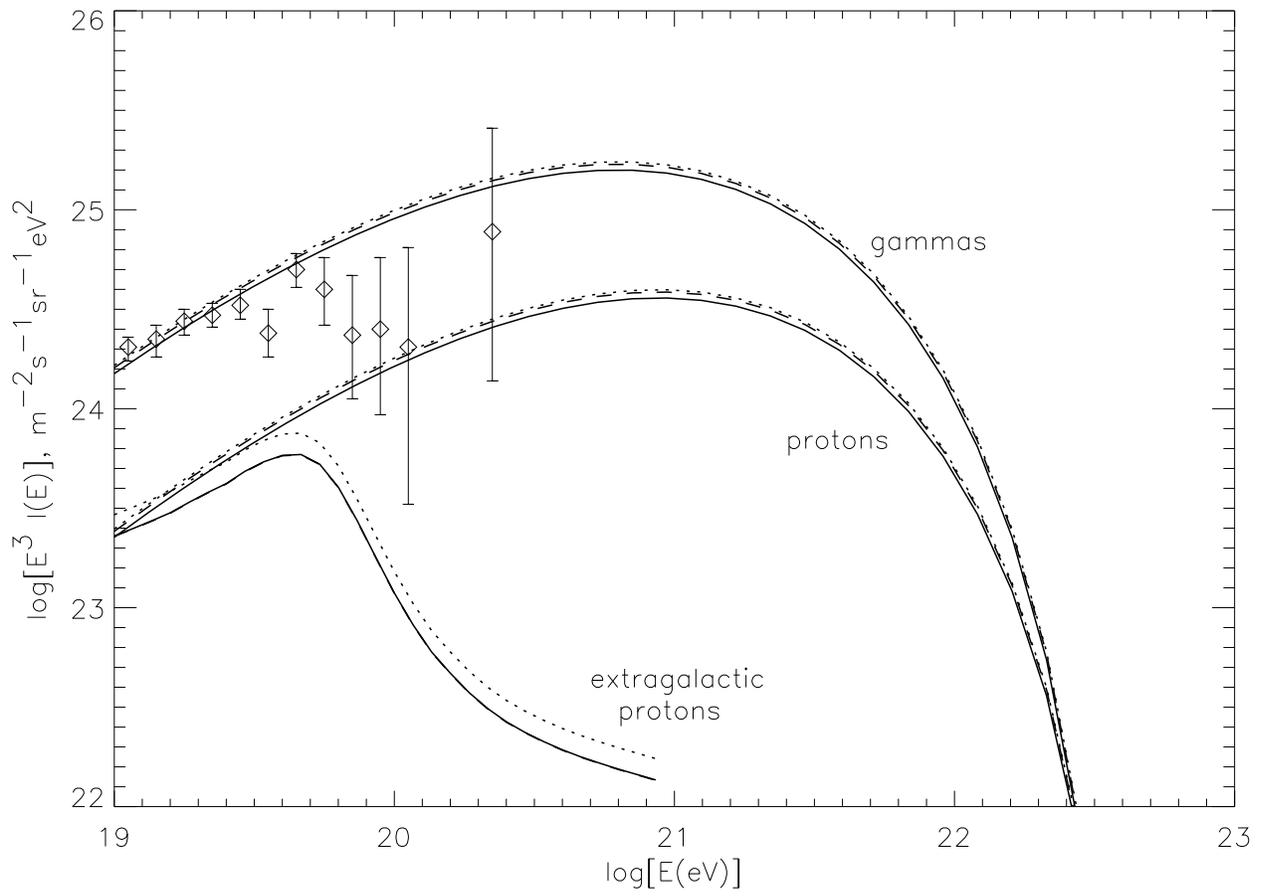
\end{document}